\documentclass[12pt,a4paper,dvips]{article}
\usepackage{a4p}
\usepackage{cite,mcite}
\usepackage{graphicx}
\usepackage{physics }
\usepackage{l3_title,ifthen}
\usepackage{rotating}
\journalname{Phys. Lett. B}

\usepackage{Lep}
\Lep{2}

\date{October 14, 1999}
\preprint{99-146}

\newlength{\capindent}
\newlength{\capwidth}
\newlength{\figwidth}
\newcommand{\icaption}[2][!*!,!]{\hspace*{\capindent}%
  \begin{minipage}{\capwidth}
    \ifthenelse{\equal{#1}{!*!,!}}%
      {\caption{#2}}%
      {\caption[#1]{#2}}
  \end{minipage}}
\def\TeV{\ifmmode {\mathrm{\ Te\kern -0.1em V}}\else
                   \textrm{Te\kern -0.1em V}\fi}%
\begin{document}
\setlength{\unitlength}{1mm}
\begin{titlepage}
\title{Search for Extra Dimensions
 in Boson and Fermion Pair Production in \boldmath{\epem} Interactions at LEP}
\author{The L3 Collaboration}
\begin{abstract}
Extra spatial dimensions are proposed by recent
theories that postulate the scale of gravity to be  of the same order
as the electroweak scale. A sizeable interaction between gravitons and
Standard Model particles is then predicted. Effects of these new
interactions in boson and fermion pair production are searched for in the
data sample collected at centre--of--mass energies above the Z pole by
the L3 detector at LEP. 
In addition, the direct production of a graviton associated with a Z
boson is investigated. No statistically significant hints for the
existence of these effects are found and lower limits in
excess of $1\TeV$ are derived on the scale of
this new theory of gravity.
\end{abstract}
\submitted

\end{titlepage}
%
%

\section{Introduction}

Contemporary collider experiments have successfully tested the 
Standard Model of electroweak
interactions (SM)~\cite{sm_glashow}  at its
characteristic distance $M_{ew}^{-1}$, where 
$M_{ew}\sim 10^{2}\GeV$ represents the electroweak scale.
The experimental study of the gravitational
force extends only down to distances  of the order of a
centimetre~\cite{expgravity}, 
 thirty
three orders of magnitude  above the  distance
$M_{Pl}^{-1}$. The Planck scale ($M_{Pl}\sim 10^{19}\GeV$) denotes the characteristic
scale of the gravitational interaction.

Deviations from the expected behaviour of gravity are expected in
theories that introduce $n$ extra spatial dimensions of
size $R$ that can be as large
as a fraction of a millimetre~\cite{arkani}. This follows from postulating a new
scale $M_S$ for the gravitational interaction, and imposing it to be
of the same order of $M_{ew}$. This Low
Scale Gravity (LSG) is related to the macroscopic expectations of
gravity in terms of the gravitational constant and hence to $M_{Pl}$
by the application of the Gauss' theorem in the extra dimensions:
\begin{equation}
M_{Pl}^2 \sim R^n M_S^{n+2}.
\end{equation}

In the LSG scenario,
spin--two gravitons couple with SM 
particles and contribute to the pair production of bosons and
fermions  in $\epem$ collisions. 
These effects were searched for previously~\cite{l3lsg}. The present analysis extends this
investigation using the same procedure
to the 176\,pb$^{-1}$ of data collected by the
L3 detector~\cite{l3_00} at LEP
in 1998 at the centre--of--mass energy $\sqrt{s}=188.7\GeV$.
Other results are described in Reference~\cite{opal1}.

LSG effects in  both boson and fermion pair production
 are described in terms of the parameter $M_S$~\cite{hewett},
interpreted as a cutoff of the theory. It appears as $1/M_S^4$ in the
LSG and SM interference terms and as $1/M_S^8$ in the pure graviton
exchange process. These terms are 
multiplied by the  factors $\lambda$ and $\lambda^2$, respectively, which
incorporate the dependence on the unknown full LSG
theory and are of order unity~\cite{hewett}. For numerical results on
 the scale $M_S$, this analysis assumes
 $\lambda = \pm 1$ to allow for different signs in the
interference between the SM and LSG contributions. Throughout this analysis the
radiative corrections to SM and LSG
processes are assumed to factorise.

LSG can also manifest itself via the direct production of a
graviton associated with either a photon or a Z boson. The first of these signatures is
investigated elsewhere~\cite{l3lsg,l3singlephoton}, while the second
is reported here for the first time.

%
%
\section{Boson Pair Production}

\begin{figure}[p]
  \begin{center}
    \begin{tabular}{cc}
      \mbox{\includegraphics[width=.5\figwidth]{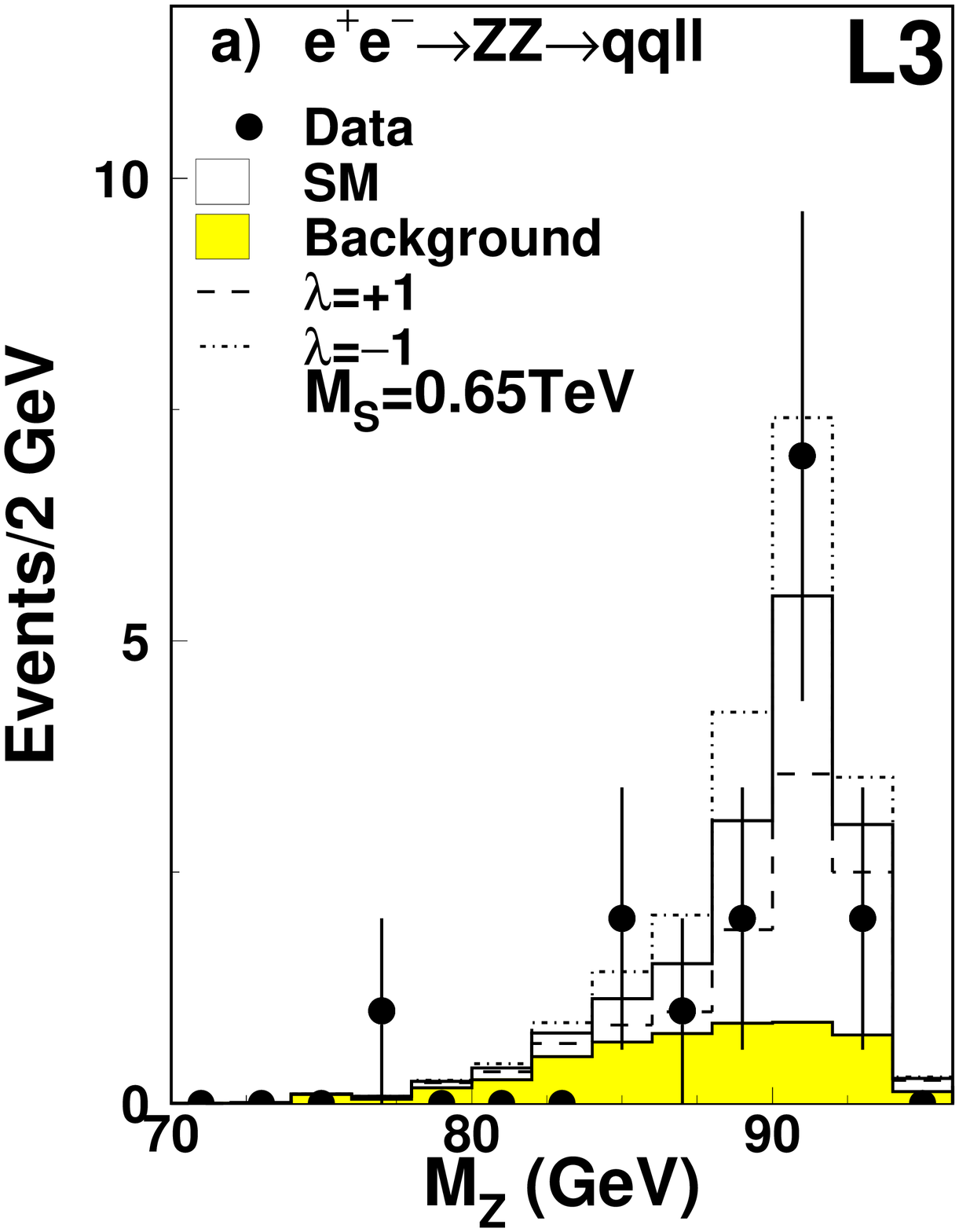}} &
      \mbox{\includegraphics[width=.5\figwidth]{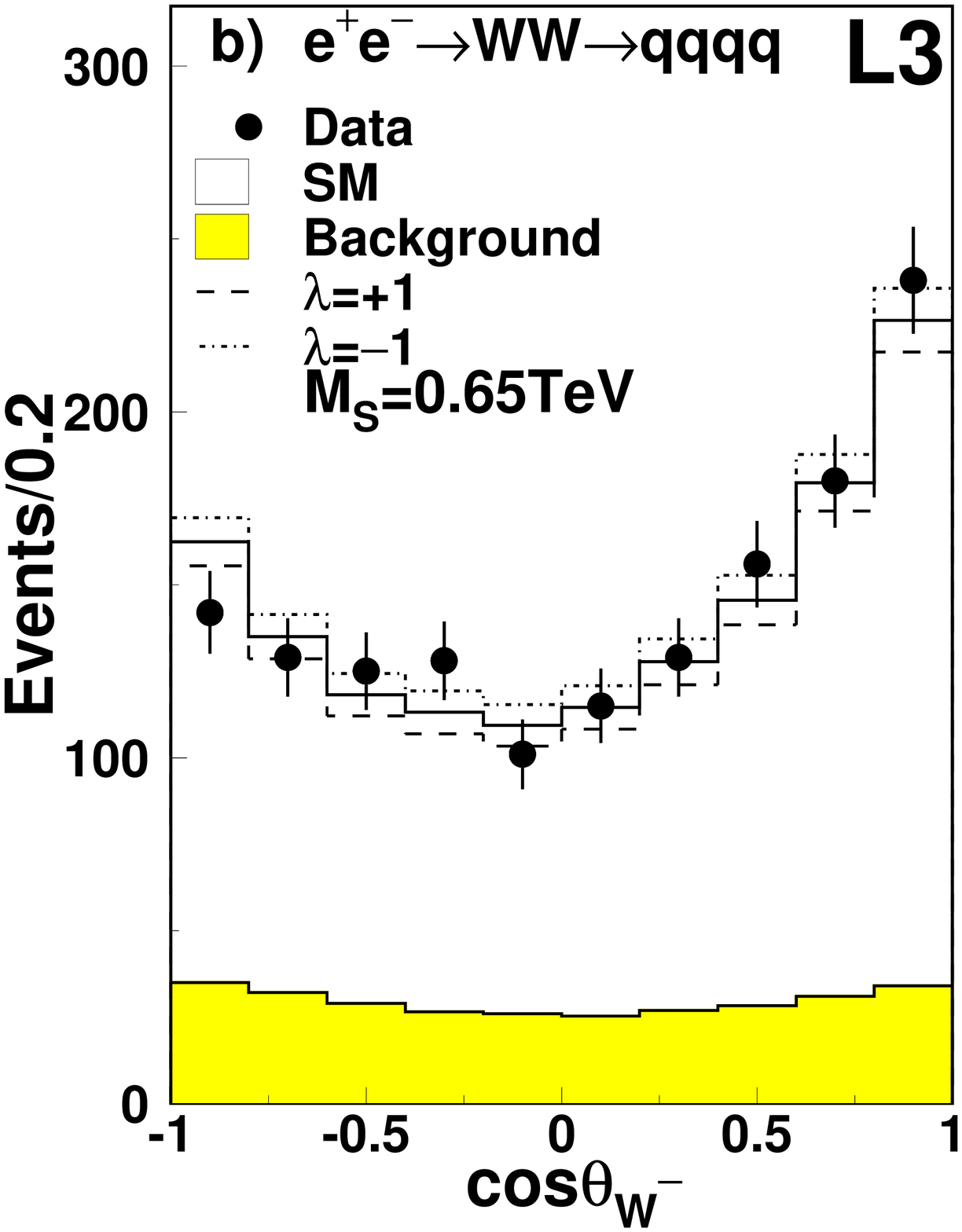}} \\
      \mbox{\includegraphics[width=.5\figwidth]{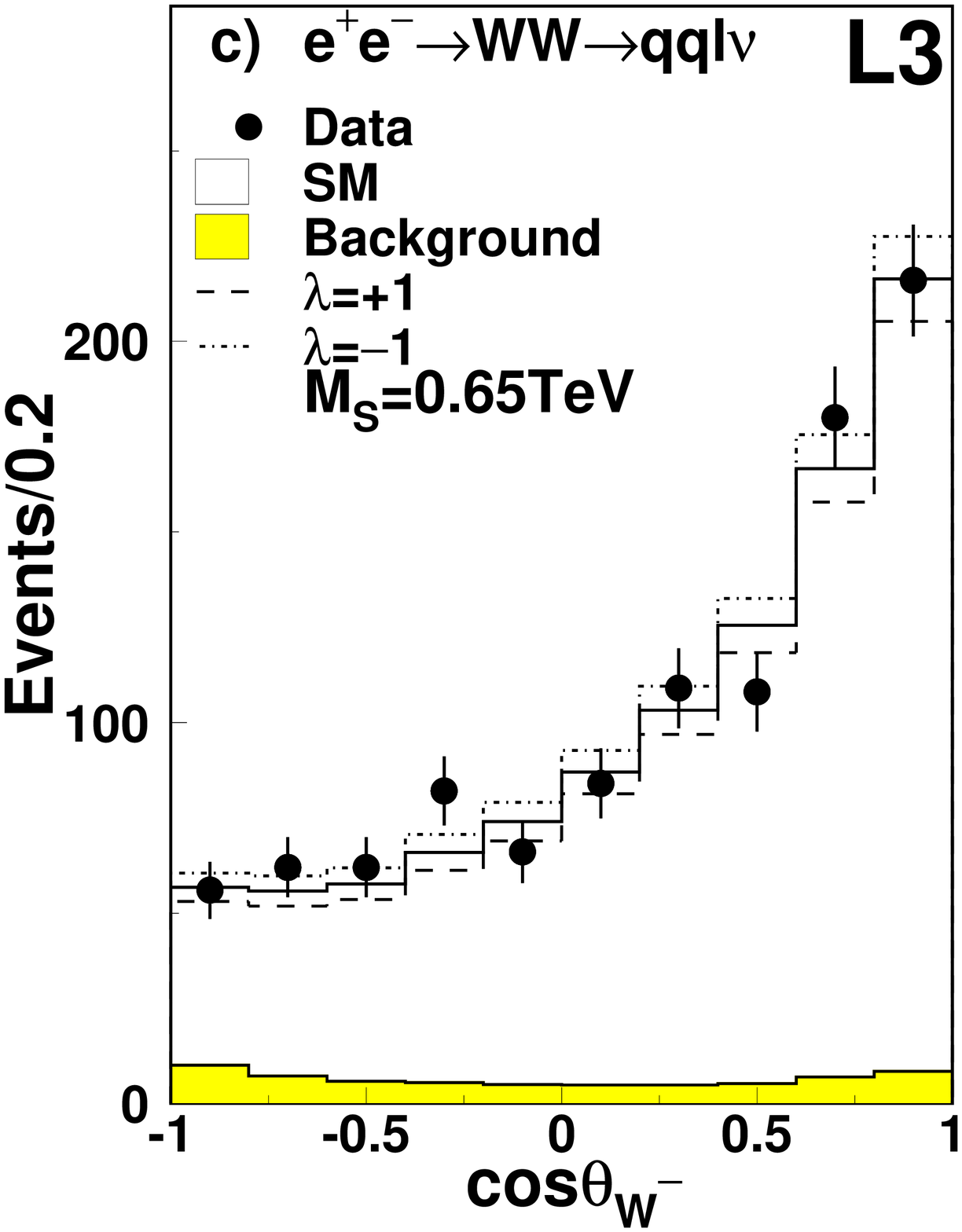}} &
      \mbox{\includegraphics[width=.5\figwidth]{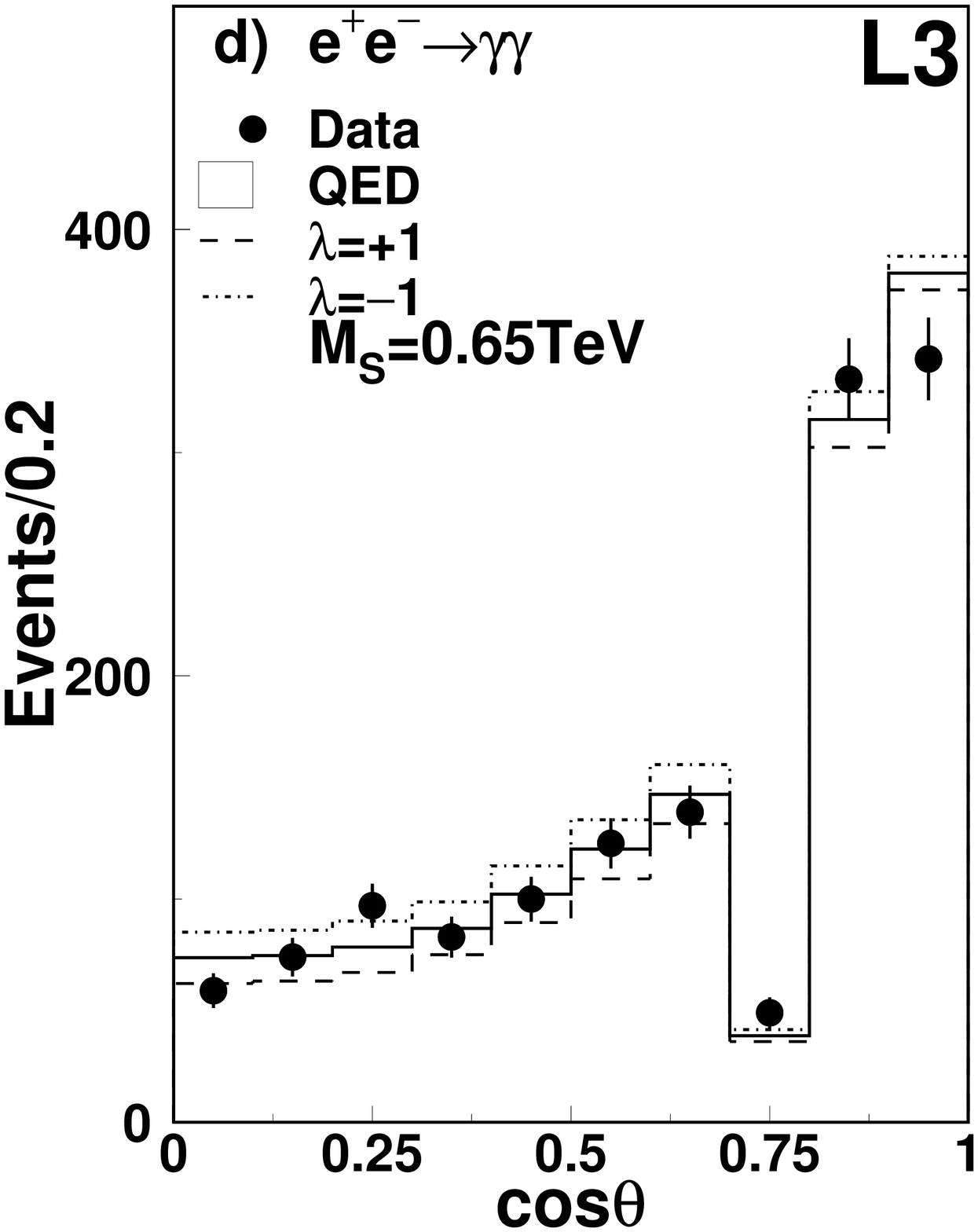}} \\
    \end{tabular}
    \icaption{(a) Reconstructed Z mass for
    $\epem\ra\Zo\Zo\ra\rm{q\overline{q}\ell^+\ell^-}$ events. Distributions of the polar angle for: (b) 
    hadronic $\epem\ra\Wp\Wm$ events, (c)  semileptonic
    $\epem\ra\Wp\Wm$ events, (d)
    $\epem\ra\gamma\gamma$ events.  Data at $188.7\GeV$, SM signal and background
    expectations are presented together with
    LSG predictions for $M_S=0.65\TeV$ and $\lambda=\pm 1$.
    \label{fig:fig1}}
  \end{center}
\end{figure}

The contribution of virtual graviton exchange to the pair production
of Z bosons affects both the total cross section and the distribution
of the Z production angle~\cite{agashe}. The same 
discriminating variables previously used to measure the ZZ cross section and
to limit a possible ZZ$\gamma$ or ZZZ vertex~\cite{l3zz189} are investigated
to search for LSG effects, namely the reconstructed Z mass of $\rm
ZZ\rightarrow q\overline{q}\ell^+\ell^-$ events, $M_{\rm Z}$, the output of a neural
network for the $\rm ZZ\rightarrow q\overline{q}q'\overline{q}'$ and
$\rm ZZ\rightarrow q\overline{q}\nu\overline{\nu}$ final states and the sum of
the visible and recoil masses in the $\rm ZZ\rightarrow
\ell^+\ell^-\ell^{'+}\ell^{'-}$ and $\rm ZZ\rightarrow
\ell^+\ell^-\nu\overline{\nu}$ channels.
ZZ events are generated
with the EXCALIBUR Monte
Carlo (MC) program~\cite{EXCALIBUR}. LSG effects are modelled by
reweighting these events with a modified
version of EXCALIBUR that includes the LSG matrix element for the ZZ final
state~\cite{agashe}. Figure~\ref{fig:fig1}a presents the distributions of
$M_{\rm Z}$ for data, SM expectations and  LSG predictions.

LSG effects in W pair
production  modify its differential cross section~\cite{agashe}, as
displayed in Figures~\ref{fig:fig1}b and~\ref{fig:fig1}c, where 
the polar angle of the emitted $\Wm$ boson is shown. Semileptonic and
hadronic decays of the W pairs are analysed~\cite{l3ww189}. Data and SM
expectations as calculated by the KORALW MC~\cite{koralw} are also
illustrated.
The inclusion of  LSG effects proceeds through the reweighting of
the 
MC events with a  modified
version of EXCALIBUR~\cite{pittauprivate}  which includes the
virtual graviton exchange matrix element~\cite{agashe} for
double--resonant processes.  A 5\%
correction is applied to account for other  diagrams contributing to the semileptonic
electron final states.

The differential cross section of the  process 
$\epem\ra\gamma\gamma$ is also sensitive to 
$s-$channel graviton exchange~\cite{giudice,agashe}.
Figure~\ref{fig:fig1}d compares the photon polar angle distribution
of the data~\cite{l3gg189} to LSG and QED predictions.

%
%
\section{Fermion Pair Production}

Contrary to boson pair production, where 
effects of 
extra dimensions are mainly expected in the total cross section, in
the case of fermion pair production distortions of the angular
distributions occur~\cite{hewett,giudice}.
Events with high effective centre--of-mass energy, $\sqrt{s'}$,  ($\sqrt{s'} >
0.85 \sqrt{s}$) are studied. 

\begin{figure}[p]
  \begin{center}
    \begin{tabular}{cc}
      \mbox{\includegraphics[width=.5\figwidth]{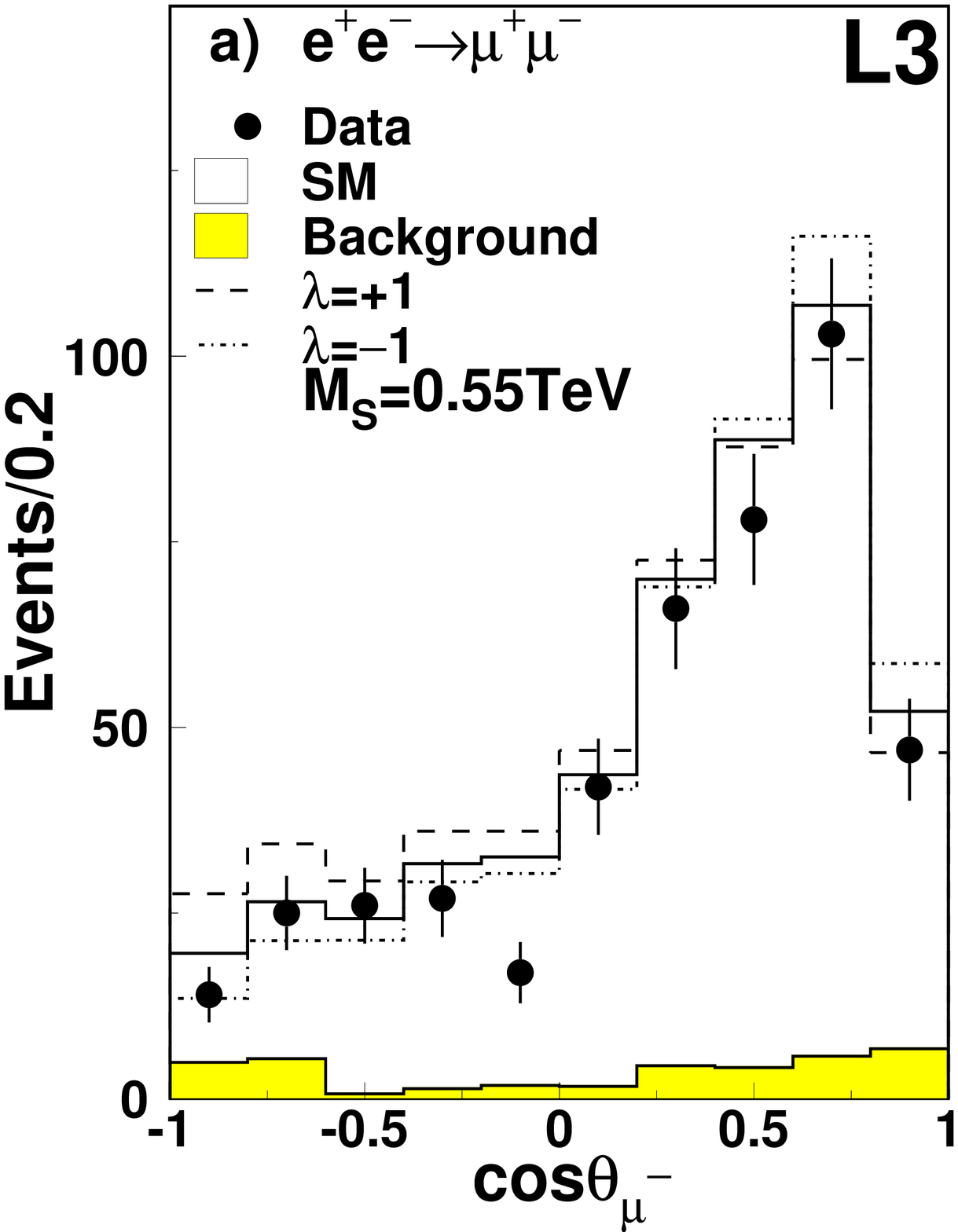}} &
      \mbox{\includegraphics[width=.5\figwidth]{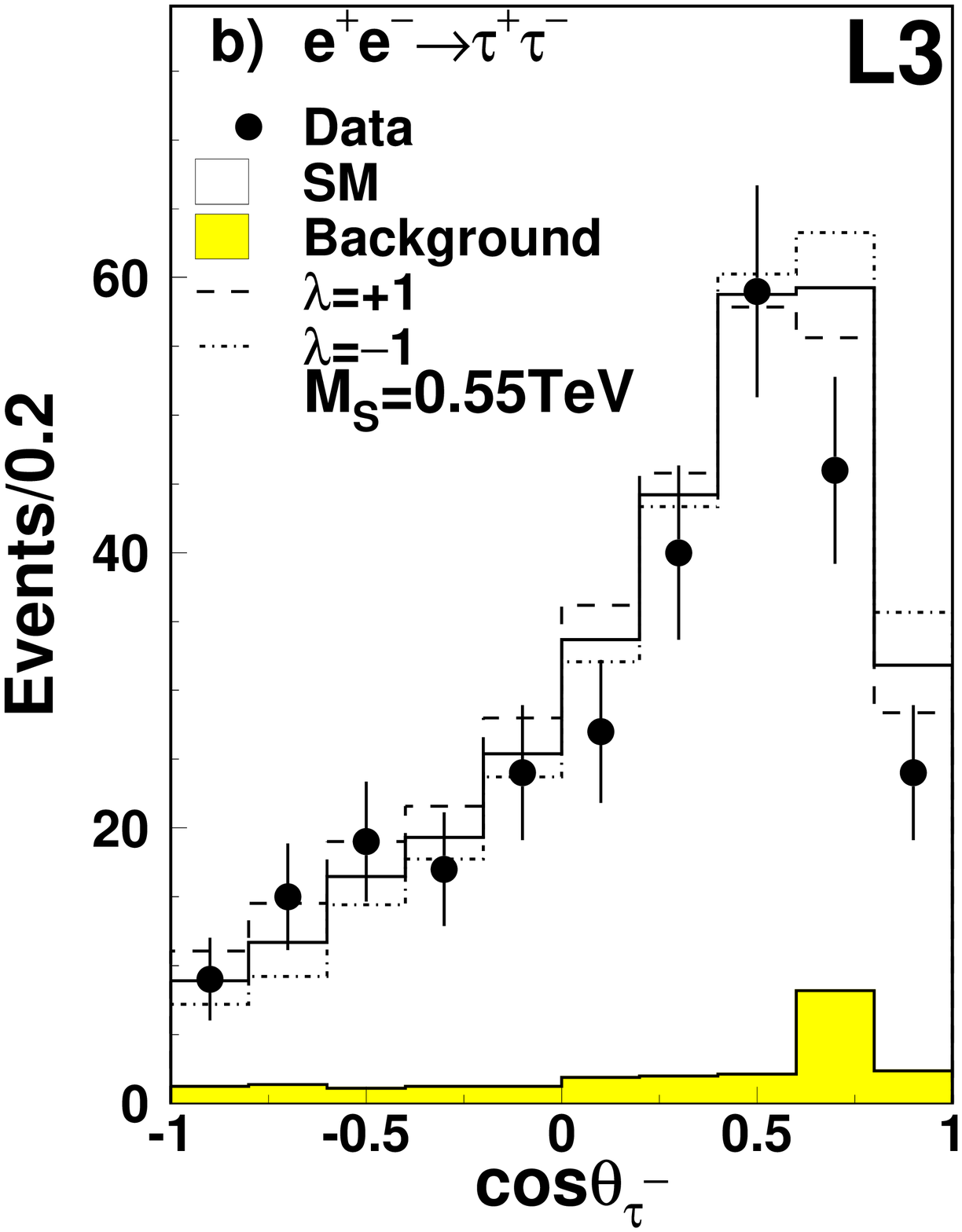}} \\
    \end{tabular}
    \icaption{Distributions of the fermion polar angle for: (a) 
      $\epem\ra\mu^+\mu^-$ and (b)  $\epem\ra\tau^+\tau^-$ processes. 
      Selected data events  at $188.7\GeV$ are shown together with SM signal and
      background expectations.  LSG predictions for the two signs of
      the interference and $M_S=0.55\TeV$ are also presented.
    \label{fig:fig2}} 
  \end{center}
\end{figure}

Figures~\ref{fig:fig2}a and~\ref{fig:fig2}b  compare the angular distributions of
muon and tau 
pairs
selected in data~\cite{ffbar189} to 
SM and LSG expectations. The SM predictions are modelled by MC events 
generated with KORALZ~\cite{KORALZ} and reweighted to the
ZFITTER~\cite{ZFITTER} differential cross sections.
The  effects of extra
dimensions are studied by reweighting these events to the LSG differential cross
sections~\cite{hewett,giudice}. 

For the $\qqbar$ final states  only the total cross
section is investigated and thus the higher sensitivity
interference term~\cite{hewett,giudice} 
vanishes. The measurement of the
cross section~\cite{ffbar189} is sensitive only to the  pure 
graviton exchange and is independent of the sign of $\lambda$.

\begin{figure}[p]
  \begin{center}
      \mbox{\includegraphics[width=\figwidth]{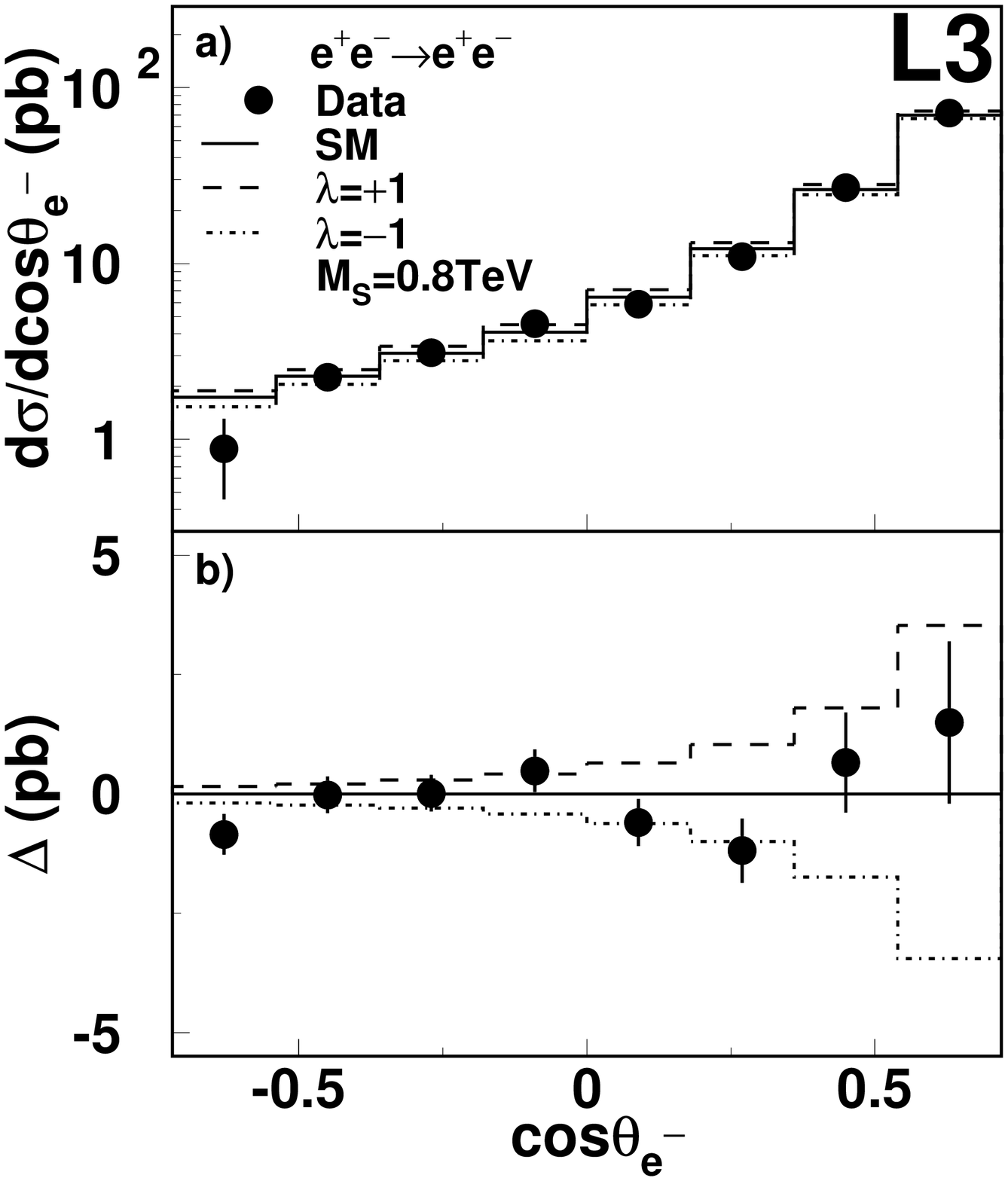}}
    \icaption{a) Measured and predicted differential cross sections for Bhabha scattering. LSG
    expectations for $\lambda=\pm 1$ and $M_S = 0.8\TeV$ are also
    shown. b) Differences $\Delta$ of the measured and LSG
    differential cross sections with respect to the SM
    prediction. Data collected  at $188.7\GeV$ are presented.
    \label{fig:fig3}} 
  \end{center}
\end{figure}

The Bhabha scattering is the channel with the highest sensitivity to
LSG effects owing to their large interference with the SM $t$--channel
diagram. Figure~\ref{fig:fig3} presents the measured differential
cross section for
events in the polar angular range of the scattered electron 
between 44$^\circ$ and 136$^\circ$. The SM predictions as obtained
from TOPAZ0~\cite{topaz0} are also shown together with the 
deviations expected from LSG~\cite{rizzo}.

%
%
\section{Results}

The distributions in Figures~\ref{fig:fig1},~\ref{fig:fig2}
and~\ref{fig:fig3}a are separately analysed 
in terms of $\lambda/M_S^4$. In addition  the  $\qqbar$ cross
section is compared with the LSG predictions in terms of
$\lambda^2/M_S^8$  and the other four
distributions describing the different decay modes of the Z boson pair
production~\cite{l3zz189} are considered as well.  Previous results from lower
$\sqrt{s}$ data~\cite{l3lsg} are included.
A likelihood is determined as a function of $\lambda/M_S^4$.  The deviation
$\delta$ of the maximum of the likelihood from the SM value of zero is measured in units
of one sigma errors and is reported in Table~1.
In all the quoted fits, the  background dependence on LSG effects is
negligible. 
No significant deviations from
the SM expectations  are found. The likelihood
functions are then integrated over the physical region for
$\lambda = +1$ and $\lambda = -1$ to yield the 95\% confidence level
(CL) limits on $M_S$, also reported in Table~1. 

The
sensitivity of the ZZ channel is comparable to that of the 
$\gamma\gamma$ and $\rm W^+W^-$ ones, despite the significantly lower
cross section. The
expected high sensitivity of the Bhabha channel is confirmed, as it dominates
the limits.

Systematic uncertainties are included in the fit. They  are calculated as the sum in quadrature of the systematic
error on the measured cross section and the theory uncertainty on its
prediction and amount to  10\% for the ZZ
channel, 4\% for WW, 1\% for $\gamma\gamma$, 2.4\% for $\mu^+\mu^-$,
3.5\% for $\tau^+\tau^-$, 1.4\% for  $\qqbar$ and 3.0\% for Bhabha scattering.
Only the limits from the $\qqbar$ and Bhabha channels are affected
by these systematic effects.

Assuming that no  higher order operators give sizeable
contributions to the LSG mediated boson and fermion pair production
and  that the meaning of the
cutoff parameter is the same for all the investigated processes, it is
possible to fit simultaneously all the boson  and 
fermion channels, and finally to combine these two results into a final
fit. No
statistically significant  extra dimensions effects are
found. The 95\% CL lower limits on $M_S$   are
listed in Table~1. They reach
$1.07\TeV$ for $\lambda = +1$ and
$0.87\TeV$ for $\lambda = -1$.

\begin{table}[ht]
  \begin{center}
    \begin{tabular}{|c|c|c|c|}
       \hline
       Process              &  $\delta$ &$M_S$\,(\TeV)& $M_S$\,(\TeV)\\
                             & &$\lambda = +1$ & $\lambda = -1$ \\
       \hline
       $\epem\ra\Zo\Zo$         & $-0.55$ & 0.77 & 0.76 \\
       $\epem\ra\Wp\Wm$         & $-1.10$ & 0.79 & 0.68 \\
       $\epem\ra\gamma\gamma$   & $-0.03$ & 0.79 & 0.80 \\
       \hline                  
       Bosons Combined          & $-0.79$ & 0.89 & 0.82 \\
       \hline                  
       $\epem\ra\mu^+\mu^-$     & $-1.12$ & 0.69 & 0.56 \\
       $\epem\ra\tau^+\tau^-$   & $+0.56$ & 0.54 & 0.58 \\
       $\epem\ra\qqbar    $     & $\pm2.30$ & 0.49 & 0.49 \\
       $\epem\ra\epem     $     & $-0.91$ & 0.98 & 0.84 \\
       \hline                                    
       Fermions Combined        &  $-1.04$ & 1.00 & 0.84 \\
       \hline                                    
       Bosons + Fermions        & $-1.30$ & 1.07 & 0.87 \\
       \hline
    \end{tabular}
    \icaption[cuts]{Lower limits at 95\,\% CL on the cutoff
    $M_S$ for different processes and values of
    $\lambda$. Deviations $\delta$ from the SM, defined in the text, are also
    given.}  
  \end{center}
\end{table}

%
%
\section{Single Z Production}

In addition to the possible LSG effects in pair production of SM
particles, the direct 
graviton (G) production associated with a Z boson is studied for the
first time, complementing the $\rm G\gamma$
search~\cite{l3lsg,l3singlephoton}. The expected cross 
section $\sigma_{\rm Z G}$ is proportional to $M_S^{-(n+2)}$~\cite{cheung} and thus
falls  rapidly with the number  of extra
dimensions $n$.
It is shown in
Table~2 for a benchmark value of $M_S = 0.5\TeV$. The symbol
$M_S$ represents  the LSG scale analogous to  
the $M_D$ parameter~\cite{giudice} investigated in the $\rm G\gamma$
channel. In the particular case of $n=2$ 
the two parameters are related by $M_S^4 = 4 M_D^4$~\cite{cheungprivate}. 
The reduced sensitivity with respect to the $\rm G\gamma$ channel follows
from the limited phase space available for graviton emission due
to the mass of the Z.

 The signature
of this process is a single Z boson in the detector as the graviton is
emitted in the extra dimensions and hence undetected. Only hadronic Z
decays are considered. A sample of 1068 unbalanced hadronic events
with missing energy pointing in the detector and a visible mass compatible
with that of the Z is
selected. The SM expectation amounts to 1096 events. The signal
efficiency is 87\%. Efficiencies
and distributions for the signal are estimated by analysing a sample
of $\epem\ra\Zo\nu\overline{\nu}$ events generated in the W fusion
process with EXCALIBUR and then reweighted to the 
Z energy and polar angle distributions of the $\rm\epem\ra Z G$
process~\cite{cheung}. 
The analysis is designed to be independent of $n$.

\begin{figure}[p]
  \begin{center}
      \mbox{\includegraphics[width=\figwidth]{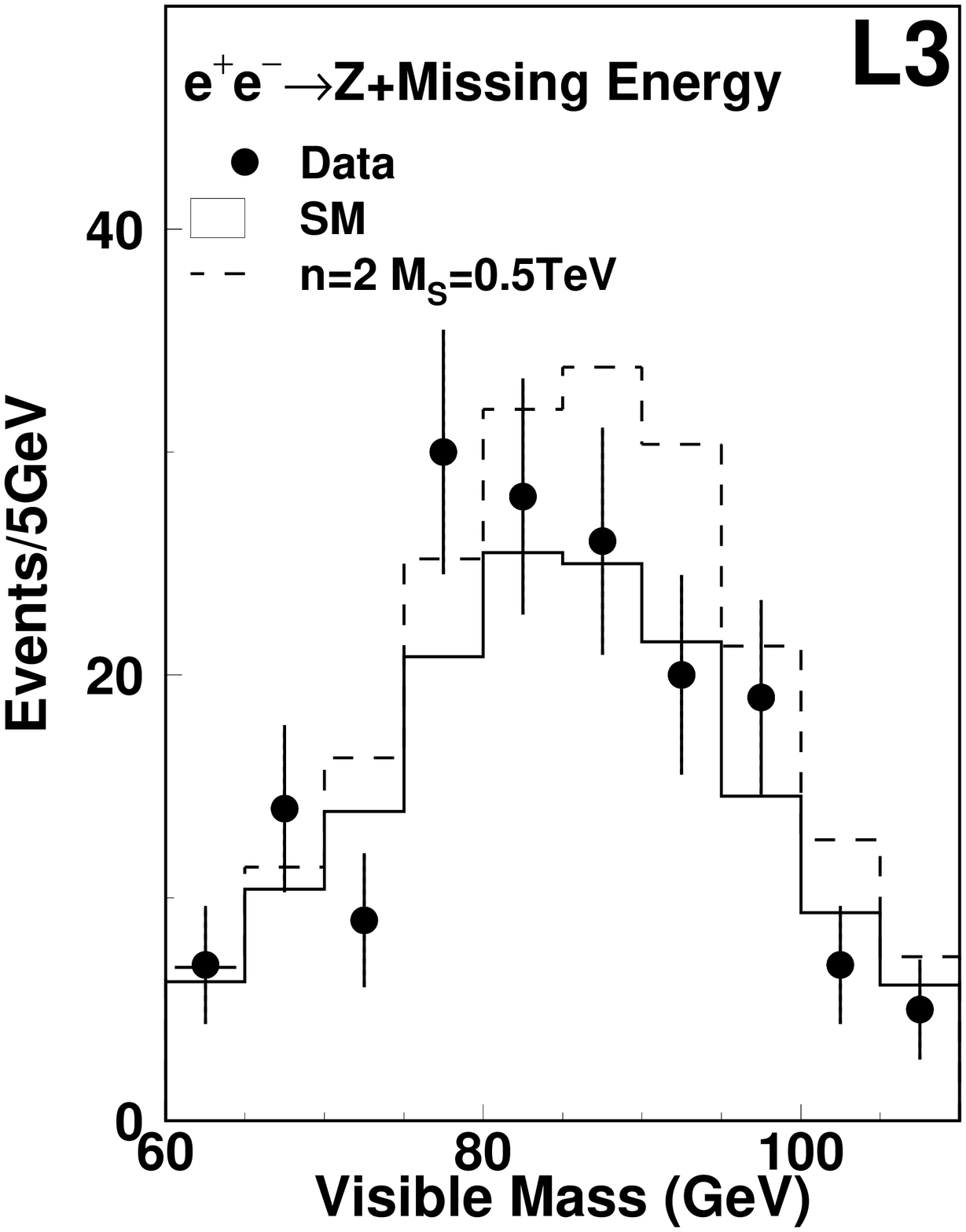}} 
    \icaption{Visible mass for $\epem\ra\Zo G$ candidate events at $188.7\GeV$
    together with SM expectations, dominated by W pair and single W production. The effect of 
    real graviton production with two extra space dimensions and
    $M_S=0.5\TeV$ is also shown.
    \label{fig:fig5}} 
  \end{center}
\end{figure}

Event--shape and jet--shape variables similar to those used in the
$\epem\ra\Zo\Zo\ra{\rm q
  \overline{q}}\nu\overline{\nu}$~\cite{l3zz189}
and hadronicly decaying single W~\cite{l3singlew} selections are used to
suppress the dominant backgrounds: radiative return to the Z,
double resonant W pair production followed by semileptonic decays into either a tau or an undetected low
angle lepton and hadronic decaying single W events. A final sample of
129 events is selected with  126 expected from SM processes with the signal 
efficiencies listed in Table~2. A fit to the visible mass distribution
of Figure~\ref{fig:fig5} yields the  95\,\% CL cross section upper limits 
in Table~2 from which the corresponding  lower limits on $M_S$ are 
extracted.

\begin{table}[ht]
  \begin{center}
    \begin{tabular}{|r|ccc|}
      \hline
      \rule{0pt}{12pt}$n$ & 2 & 3 & 4    \\ 
      \hline
      \rule{0pt}{12pt}$\sigma_{\rm Z G}$ (pb) & 0.64 & 0.081 &
      0.011  \\ 
      \hline
      \rule{0pt}{12pt} $\varepsilon$  & 0.56 & 0.56 & 0.55  \\
      \rule{0pt}{12pt}$\sigma_{\rm Z G}^{\rm lim}$ (pb) & 0.29 & 0.30 &
      0.30  \\ 
      \rule{0pt}{12pt}$M_S$ (TeV) & 0.60 & 0.38 & 0.29   \\ 
      \hline
    \end{tabular}
    \icaption[cuts]{Expected cross sections $\sigma_{\rm Z G}$ for the graviton
    plus Z signal $(M_S = 0.5\TeV)$, detection efficiency
    $\varepsilon$, upper limit at 95\,\% CL $\sigma_{\rm Z G}^{\rm lim}$ on  the cross
    section and lower limit on the
    scale $M_S$ as a function of  
    the number of extra dimensions $n$.}
  \end{center}
\end{table}

In conclusion no evidence for extra dimensions is found and limits in excess of $1\TeV$ are set on the scale of LSG.

%
%
\section*{Acknowledgements}

We thank Kingman Cheung for useful discussions on the
associated $\Zo$ and graviton production.
We wish to express our gratitude to the CERN accelerator divisions for the
superb performance and the continuous and successful upgrade of the
LEP machine.  
We acknowledge the contributions of the engineers  and technicians who
have participated in the construction and maintenance of this experiment.

%
%

\bibliographystyle{l3stylem}

%
%

\newpage
\typeout{   }     
\typeout{Using author list for paper 192 -?}
\typeout{$Modified: Wed Oct 13 08:43:45 1999 by clare $}
\typeout{!!!!  This should only be used with document option a4p!!!!}
\typeout{   }
%
%
%
%
%
%

\newcount\tutecount  \tutecount=0
\def\tutenum#1{\global\advance\tutecount by 1 \xdef#1{\the\tutecount}}
\def\tute#1{$^{#1}$}
\tutenum\aachen            
\tutenum\nikhef            
\tutenum\mich              
\tutenum\lapp              
\tutenum\basel             
\tutenum\lsu               
\tutenum\beijing           
\tutenum\berlin            
\tutenum\bologna           
\tutenum\tata              
\tutenum\ne                
\tutenum\bucharest         
\tutenum\budapest          
\tutenum\mit               
\tutenum\debrecen          
\tutenum\florence          
\tutenum\cern              
\tutenum\wl                
\tutenum\geneva            
\tutenum\hefei             
\tutenum\seft              
\tutenum\lausanne          
\tutenum\lecce             
\tutenum\lyon              
\tutenum\madrid            
\tutenum\milan             
\tutenum\moscow            
\tutenum\naples            
\tutenum\cyprus            
\tutenum\nymegen           
\tutenum\caltech           
\tutenum\perugia           
\tutenum\cmu               
\tutenum\prince            
\tutenum\rome              
\tutenum\peters            
\tutenum\salerno           
\tutenum\ucsd              
\tutenum\santiago          
\tutenum\sofia             
\tutenum\korea             
\tutenum\alabama           
\tutenum\utrecht           
\tutenum\purdue            
\tutenum\psinst            
\tutenum\zeuthen           
\tutenum\eth               
\tutenum\hamburg           
\tutenum\taiwan            
\tutenum\tsinghua          
{
\parskip=0pt
\noindent
{\bf The L3 Collaboration:}
\ifx\selectfont\undefined
 \baselineskip=10.8pt
 \baselineskip\baselinestretch\baselineskip
 \normalbaselineskip\baselineskip
 \ixpt
\else
 \fontsize{9}{10.8pt}\selectfont
\fi
\medskip
\tolerance=10000
\hbadness=5000
\raggedright
\hsize=162truemm\hoffset=0mm
\def\r{\rlap,}
\noindent

M.Acciarri\r\tute\milan\
P.Achard\r\tute\geneva\ 
O.Adriani\r\tute{\florence}\ 
M.Aguilar-Benitez\r\tute\madrid\ 
J.Alcaraz\r\tute\madrid\ 
G.Alemanni\r\tute\lausanne\
J.Allaby\r\tute\cern\
A.Aloisio\r\tute\naples\ 
M.G.Alviggi\r\tute\naples\
G.Ambrosi\r\tute\geneva\
% ...
\newpage
\begin{list}{A}{\itemsep=0pt plus 0pt minus 0pt\parsep=0pt plus 0pt minus 0pt
                \topsep=0pt plus 0pt minus 0pt}
\item[\aachen]
 I. Physikalisches Institut, RWTH, D-52056 Aachen, FRG$^{\S}$\\
 III. Physikalisches Institut, RWTH, D-52056 Aachen, FRG$^{\S}$
\item[\nikhef] National Institute for High Energy Physics, NIKHEF, 
     and University of Amsterdam, NL-1009 DB Amsterdam, The Netherlands
\item[\mich] University of Michigan, Ann Arbor, MI 48109, USA
\item[\lapp] Laboratoire d'Annecy-le-Vieux de Physique des Particules, 
     LAPP,IN2P3-CNRS, BP 110, F-74941 Annecy-le-Vieux CEDEX, France
\item[\basel] Institute of Physics, University of Basel, CH-4056 Basel,
     Switzerland
\item[\lsu] Louisiana State University, Baton Rouge, LA 70803, USA
\item[\beijing] Institute of High Energy Physics, IHEP, 
  100039 Beijing, China$^{\triangle}$ 
\item[\berlin] Humboldt University, D-10099 Berlin, FRG$^{\S}$
\item[\bologna] University of Bologna and INFN-Sezione di Bologna, 
     I-40126 Bologna, Italy
\item[\tata] Tata Institute of Fundamental Research, Bombay 400 005, India
\item[\ne] Northeastern University, Boston, MA 02115, USA
\item[\bucharest] Institute of Atomic Physics and University of Bucharest,
     R-76900 Bucharest, Romania
\item[\budapest] Central Research Institute for Physics of the 
     Hungarian Academy of Sciences, H-1525 Budapest 114, Hungary$^{\ddag}$
\item[\mit] Massachusetts Institute of Technology, Cambridge, MA 02139, USA
\item[\debrecen] KLTE-ATOMKI, H-4010 Debrecen, Hungary$^\P$
\item[\florence] INFN Sezione di Firenze and University of Florence, 
     I-50125 Florence, Italy
\item[\cern] European Laboratory for Particle Physics, CERN, 
     CH-1211 Geneva 23, Switzerland
\item[\wl] World Laboratory, FBLJA  Project, CH-1211 Geneva 23, Switzerland
\item[\geneva] University of Geneva, CH-1211 Geneva 4, Switzerland
\item[\hefei] Chinese University of Science and Technology, USTC,
      Hefei, Anhui 230 029, China$^{\triangle}$
\item[\seft] SEFT, Research Institute for High Energy Physics, P.O. Box 9,
      SF-00014 Helsinki, Finland
\item[\lausanne] University of Lausanne, CH-1015 Lausanne, Switzerland
\item[\lecce] INFN-Sezione di Lecce and Universit\'a Degli Studi di Lecce,
     I-73100 Lecce, Italy
\item[\lyon] Institut de Physique Nucl\'eaire de Lyon, 
     IN2P3-CNRS,Universit\'e Claude Bernard, 
     F-69622 Villeurbanne, France
\item[\madrid] Centro de Investigaciones Energ{\'e}ticas, 
     Medioambientales y Tecnolog{\'\i}cas, CIEMAT, E-28040 Madrid,
     Spain${\flat}$ 
\item[\milan] INFN-Sezione di Milano, I-20133 Milan, Italy
\item[\moscow] Institute of Theoretical and Experimental Physics, ITEP, 
     Moscow, Russia
\item[\naples] INFN-Sezione di Napoli and University of Naples, 
     I-80125 Naples, Italy
\item[\cyprus] Department of Natural Sciences, University of Cyprus,
     Nicosia, Cyprus
\item[\nymegen] University of Nijmegen and NIKHEF, 
     NL-6525 ED Nijmegen, The Netherlands
\item[\caltech] California Institute of Technology, Pasadena, CA 91125, USA
\item[\perugia] INFN-Sezione di Perugia and Universit\'a Degli 
     Studi di Perugia, I-06100 Perugia, Italy   
\item[\cmu] Carnegie Mellon University, Pittsburgh, PA 15213, USA
\item[\prince] Princeton University, Princeton, NJ 08544, USA
\item[\rome] INFN-Sezione di Roma and University of Rome, ``La Sapienza",
     I-00185 Rome, Italy
\item[\peters] Nuclear Physics Institute, St. Petersburg, Russia
\item[\salerno] University and INFN, Salerno, I-84100 Salerno, Italy
\item[\ucsd] University of California, San Diego, CA 92093, USA
\item[\santiago] Dept. de Fisica de Particulas Elementales, Univ. de Santiago,
     E-15706 Santiago de Compostela, Spain
\item[\sofia] Bulgarian Academy of Sciences, Central Lab.~of 
     Mechatronics and Instrumentation, BU-1113 Sofia, Bulgaria
\item[\korea] Center for High Energy Physics, Adv.~Inst.~of Sciences
     and Technology, 305-701 Taejon,~Republic~of~{Korea}
\item[\alabama] University of Alabama, Tuscaloosa, AL 35486, USA
\item[\utrecht] Utrecht University and NIKHEF, NL-3584 CB Utrecht, 
     The Netherlands
\item[\purdue] Purdue University, West Lafayette, IN 47907, USA
\item[\psinst] Paul Scherrer Institut, PSI, CH-5232 Villigen, Switzerland
\item[\zeuthen] DESY, D-15738 Zeuthen, 
     FRG
\item[\eth] Eidgen\"ossische Technische Hochschule, ETH Z\"urich,
     CH-8093 Z\"urich, Switzerland
\item[\hamburg] University of Hamburg, D-22761 Hamburg, FRG
\item[\taiwan] National Central University, Chung-Li, Taiwan, China
\item[\tsinghua] Department of Physics, National Tsing Hua University,
      Taiwan, China
\item[\S]  Supported by the German Bundesministerium 
        f\"ur Bildung, Wissenschaft, Forschung und Technologie
\item[\ddag] Supported by the Hungarian OTKA fund under contract
numbers T019181, F023259 and T024011.
\item[\P] Also supported by the Hungarian OTKA fund under contract
  numbers T22238 and T026178.
\item[$\flat$] Supported also by the Comisi\'on Interministerial de Ciencia y 
        Tecnolog{\'\i}a.
\item[$\sharp$] Also supported by CONICET and Universidad Nacional de La Plata,
        CC 67, 1900 La Plata, Argentina.
\item[$\diamondsuit$] Also supported by Panjab University, Chandigarh-160014, 
        India.
\item[$\triangle$] Supported by the National Natural Science
  Foundation of China.
\item[\dag] Deceased.
\end{list}
}
\vfill






\newpage

%
%

\end{document}